\documentclass[aps,reprint,superscriptaddress]{revtex4-1}

\usepackage{graphicx}

\begin{document}

\title{Wetting and phase separation in soft adhesion}

\author{K. E. Jensen}
\affiliation{Yale University, New Haven, CT, 06511, USA}

\author{R. Sarfati}
\affiliation{Yale University, New Haven, CT, 06511, USA}

\author{R. W. Style}
\affiliation{Mathematical Institute, University of Oxford, Oxford, OX1 3LB, UK}

\author{R. Boltyanskiy}
\affiliation{Yale University, New Haven, CT, 06511, USA}

\author{A. Chakrabarti}
\affiliation{Department of Chemical Engineering, Lehigh University, Bethlehem, PA, 18015, USA}

\author{M. K. Chaudhury}
\affiliation{Department of Chemical Engineering, Lehigh University, Bethlehem, PA, 18015, USA}

\author{E. R. Dufresne}
\affiliation{Yale University, New Haven, CT, 06511, USA}

\date{\today}

\begin{abstract}

In the classic theory of solid adhesion, surface energy drives deformation to increase contact area while bulk elasticity opposes it.
Recently, solid surface stress has been shown also to play an important role in opposing deformation of soft materials.
This suggests that the contact line in soft adhesion should mimic that of a liquid droplet, with a contact angle determined by surface tensions.
Consistent with this hypothesis, we observe a contact angle of a soft silicone substrate on rigid silica spheres that depends on the surface functionalization but not the sphere size. 
However, to satisfy this wetting condition without  a divergent elastic stress, the gel separates from its solvent  near the contact line.
This creates a four-phase contact zone with two additional contact lines hidden below the surface of the substrate.
While the geometries of these contact lines are independent of the size of the sphere,  the volume of the phase-separated region is not, but rather depends on the indentation volume.
These results indicate that theories of adhesion of soft gels need to account for both the compressibility of the gel network and a non-zero surface stress between the gel and its solvent.

\end{abstract}


\maketitle

\section{Introduction}
Solid surfaces stick together to minimize their total surface energy. 
However, if the surfaces are not flat, they must conform to one another to make adhesive contact.
Whether or not this contact can be made, and how effectively it can be made, are crucial questions in the study and development of solid adhesive materials \cite{JohnsonBook1987, CretonPapon2003}.
These questions have wide-ranging technological consequence.
With applications ranging from construction to medicine, large-scale manufacturing to everyday sticky stuff, adhesive materials are ubiquitous in daily life.
However, much remains unknown about the mechanics of solid adhesion, especially when the solids are very compliant \cite{ShullCrosby1998, StyleNatComm2013, Pastewka2014}. 
This limits our understanding and development of anything that relies on the mechanics of soft contact, including pressure-sensitive adhesives \cite{CrosbyShull1999, Creton2003}, rubber friction \cite{Persson2001}, materials for soft robotics \cite{Kao2004, Kim2007, Martinez2013, Kim2013}, and the mechanical characterization of soft materials, including living cells \cite{Mowery1997, vanVliet2003,Suresh2007, Lim2008, BrochardWyart2012}.

Adhesion is favorable whenever the adhesion energy, $W= \gamma_1 + \gamma_2 - \gamma_{12}$, is positive,
where $\gamma_1$ and $\gamma_2$ are the surface energies of the free surfaces and $\gamma_{12}$ is the interfacial energy in contact.
When $W>0$, the solids are driven to deform spontaneously to increase their area of contact, but at the cost of incurring elastic strain. 
The foundational and widely-applied Johnson-Kendall-Roberts (JKR) theory of contact mechanics \cite{JKR1971, Maugis1995} was the first to describe this competition between adhesion and elasticity. 
However, it was recently shown that the JKR theory does not accurately describe adhesive contact with soft materials because it does not account for an additional penalty against deformation due to solid surface stress, $\Upsilon$ \cite{StyleNatComm2013}.
In general, surface stresses overwhelm elastic response when the characteristic length scale of deformation is less than an elastocapillary length, $L$, given by the ratio of the surface stress to Young's modulus, $L = \Upsilon/E$ \cite{Long1996, Jerison2011,Jagota2012,Paretkar2014,Mora2010}. 
This has an important implication for soft adhesion \cite{StyleNatComm2013, Xu2014,  Liu2015, Cao2014, Salez2013, Hui2015}: 
that the geometry of the contact line between a rigid indenter and a soft substrate should be determined by a balance of surface stresses and surface energies, just as the Young-Dupr\'{e} relation sets the contact angle of a fluid on a rigid solid \cite{deGennes2004}.
However, the structure of the contact zone in soft adhesion has not been examined experimentally.

In this article, we directly image the contact zone of rigid spheres adhered to compliant gels.
Consistent with the dominance of surfaces stresses over bulk elastic stresses,  we find that the surface of the soft substrate meets each sphere with a constant contact angle that depends on the sphere's surface functionalization but not its size. 
To satisfy this wetting condition while avoiding a divergent elastic stress, the gel and its solvent phase-separate near the contact line. 
The resulting four-phase contact zone includes two additional contact lines hidden below the liquid surface.
The geometries of all three contact lines are independent of the size of the sphere and depend on the relevant surface energies and surface stresses.
Surprisingly, these results demonstrate a finite surface stress between the gel and its solvent.
The volume of the phase-separated contact zone depends on the indentation volume and the compressibility of the gel's elastic network.

\section{Structure of the adhesive contact line}
We study the contact between rigid glass spheres and compliant silicone gels.  
Glass spheres ranging in radius from 7 to 32 $\mu$m (Polysciences, 07668) are used as-received or surface functionalized with 1H,1H,2H,2H-Perfluorooctyl-trichlorosilane (Sigma-Aldrich, 448931).
We prepare silicone gels by mixing liquid (1 Pa$\cdot$s) divinyl-terminated polydimethylsiloxane (PDMS) (Gelest, DMS-V31) with a chemical crosslinker (Gelest, HMS-301) and catalyst (Gelest, SIP6831.2).  
The silicone mixture is degassed in vacuum, put into the appropriate experimental geometry, and cured at 68$^\circ$C for 12-14 hours. 
The resulting gel is an elastic network of cross-linked polymers swollen with free liquid of the same polymer. 
The fraction of liquid  PDMS in these gels is 62\% by weight, measured by solvent extraction.
The gel has a shear modulus of $G' = 1.9$ kPa, measured by bulk rheology.
The Poisson ratio of the gel's elastic network is $\nu = 0.48$, measured using a compression test in the rheometer as described in Ref. \cite{TFMpaper}.
As this is an isotropic, elastic material, this gives a Young modulus $E= 5.6$ kPa and a bulk modulus $K = 53$ kPa. 

\begin{figure}[h!]
\includegraphics[width = 0.48 \textwidth]{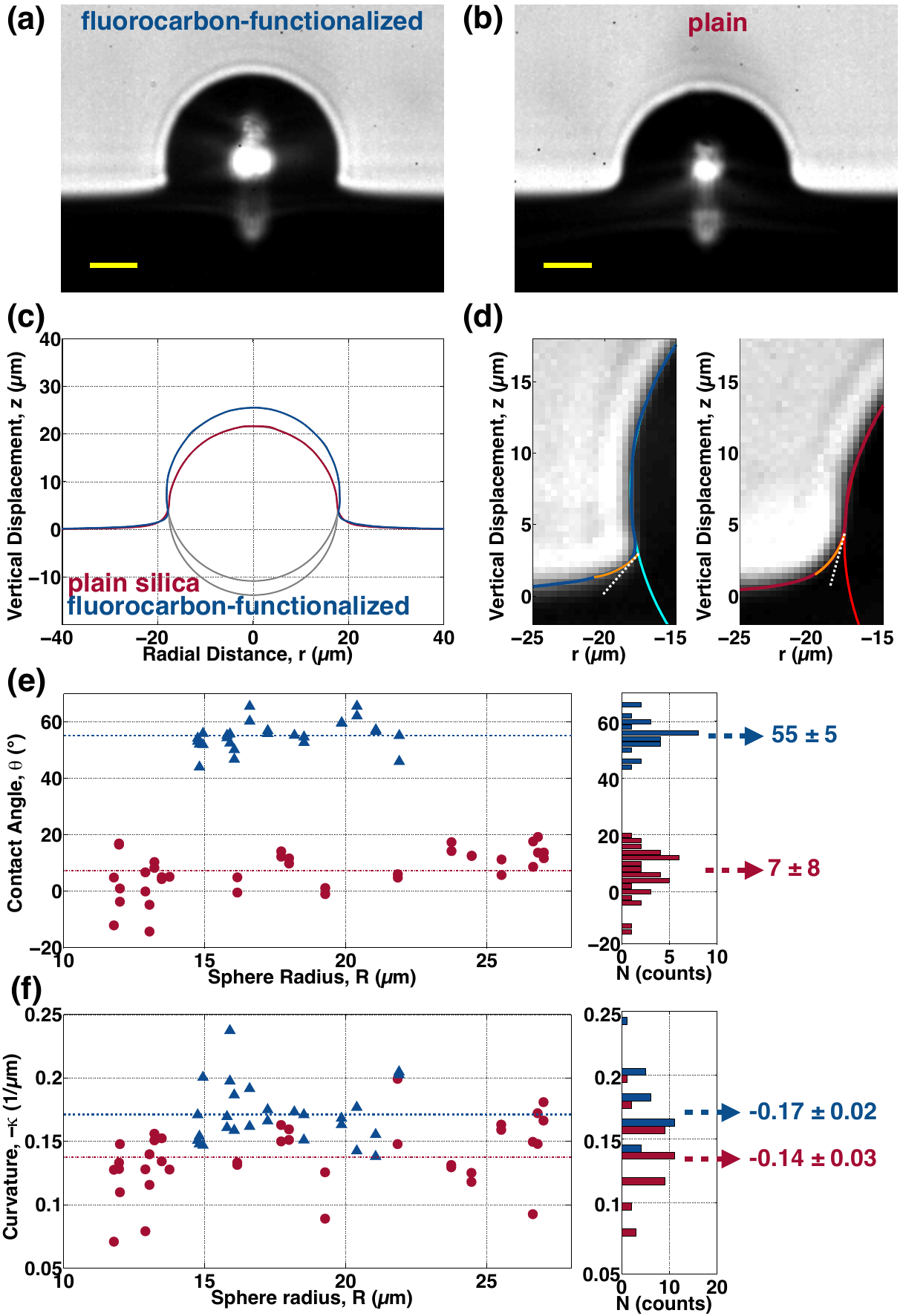}
\caption{\label{brightfield} Contact angle measurements.
\emph{(a-b)} Side views of (a) an 18.2-$\mu$m-radius fluorocarbon-functionalized silica sphere and (b) a 17.7-$\mu$m-radius plain silica sphere, each adhered to an $E = 5.6$ kPa silicone gel. Scale bars are 10 $\mu$m.
\emph{(c)} Mapped profiles of the spheres in (a-b), overlaid, with fit circles drawn to outline each sphere's position. 
The undeformed plane far from the adhered particles defines $z=0$.
\emph{(d)} Close-up of the profiles in (c) superimposed on the raw data, focusing on the approach to contact. The constant curvature fits are overlaid as orange curves, as well as straight dashed lines indicating the measured contact angles.
\emph{(e-f)} Measured contact angle, $\theta$, and measured curvature, $-\kappa$, respectively, versus sphere radius for both the fluorocarbon-functionalized (blue triangles) and plain silica (red circles) spheres. 
Dashed lines indicate the mean values. 
Histograms of the measurements are shown at right, with mean and standard deviation indicated.}
\end{figure}

We directly image the geometry of the contact between the gel and sphere using optical microscopy.
To prepare the gel substrates, we deposit a $\sim$300-$\mu$m-thick layer of PDMS along the mm-wide edge of a standard microscope slide.
The silicone surface is flat parallel to the edge of the slide and has a radius of curvature $\sim$700 $\mu$m in the orthogonal direction. 
We distribute silica spheres sparsely on the surface of the gel and image only those spheres that adhere at the peak of the gel. 
Using an inverted optical microscope, we illuminate the sample with a low N.A. condenser and image using a 40$\times$ (N.A. 0.60) air objective.
Example images for fluorocarbon-functionalized and plain silica spheres having radii of about 18 $\mu$m are shown in Figure \ref{brightfield}(a) and \ref{brightfield}(b), respectively.
In all cases, the rigid particles spontaneously indent into the gel as they adhere.
Plain silica spheres indent more deeply than fluorocarbon-functionalized spheres of the same size.

To test whether surface stresses dominate over elasticity at the contact line, we measure the contact angle between the free surface of the gel and the sphere.
Starting with the raw image data, we map the position of the dark edge in the images with 100-nm-resolution using edge detection in MATLAB.
Example profiles for fluorocarbon-functionalized (blue points) and plain silica (red points) spheres are shown in Figure \ref{brightfield}(c).
We fit the central region of the profile with a circle to determine the position and radius of the sphere, indicated by the gray lines in Figure \ref{brightfield}(c).

The approach to contact is qualitatively different for the two types of spheres;
the substrate meets the plain spheres at a much shallower angle than the fluorocarbon-functionalized ones.
We fit the substrate surface profile near the contact line to a surface of constant total curvature, which is the shape expected when surface stresses completely overwhelm elastic effects \cite{deGennes2004}.
Fit results for the profiles shown in Figure \ref{brightfield}(c) are plotted in Figure \ref{brightfield}(d), zoomed in close to the contact line on one side.
Note that we do not fit to the profile data within one micron of the contact line, since diffraction tends to round off sharp corners.
The resulting contact angles and curvatures are plotted as a function of sphere size for both fluorocarbon-functionalized and plain spheres ranging in radius from 12-27 $\mu$m in Figure \ref{brightfield}(e) and (f).

The contact angle of the substrate on the sphere is independent of the sphere size, but depends on the sphere's surface functionalization.
The gel establishes a contact angle of $\theta = 55 \pm 5^\circ$ with the fluorocarbon-functionalized spheres, and $\theta = 7 \pm 8^\circ$ with the plain spheres.
We also see no size-dependence of the curvature of the gel near the contact line, and little difference with surface functionalization: $\kappa_{plain} = -0.14 \pm 0.03$ $\mu$m$^{-1}$ and $\kappa_{fc} = -0.17 \pm 0.02$ $\mu$m$^{-1}$.
Assuming that the surface tension of the solid is close to that of the liquid, 20 mN/m, these constant curvature values are comparable to the inverse of the elastocapillary length of the substrate $E/\Upsilon=0.28$ $\mu$m$^{-1}$.

For comparison, we measure the contact angle between the spheres and uncured PDMS liquid.
In this case, the contact angles should be set by the surface energies through the classic Young-Dupr\'e relation.
We find that the plain silica spheres are completely engulfed by the silicone liquid, corresponding to a contact angle $\theta =0$.
On the fluorocarbon-functionalized spheres, the uncured liquid makes a contact angle $\theta = 54 \pm 4^\circ$.

The contact angles made by the silicone gel on the spheres are the same as the contact angles made by the silicone liquid.
This suggests that the Young-Dupr\'{e} relation governs the contact line of a soft adhesive. 
However, achieving the contact angle prescribed by Young-Dupr\'e presents a serious difficulty for the gel's elastic network, especially during contact with surfaces that demand total wetting.
As the contact angle of the gel approaches zero, the tensile strain on the elastic network diverges. 
How does the gel satisfy the wetting condition without creating an elastic singularity?

\section{Deformation of the elastic network}
\begin{figure}[h]
\includegraphics[width = 0.48 \textwidth]{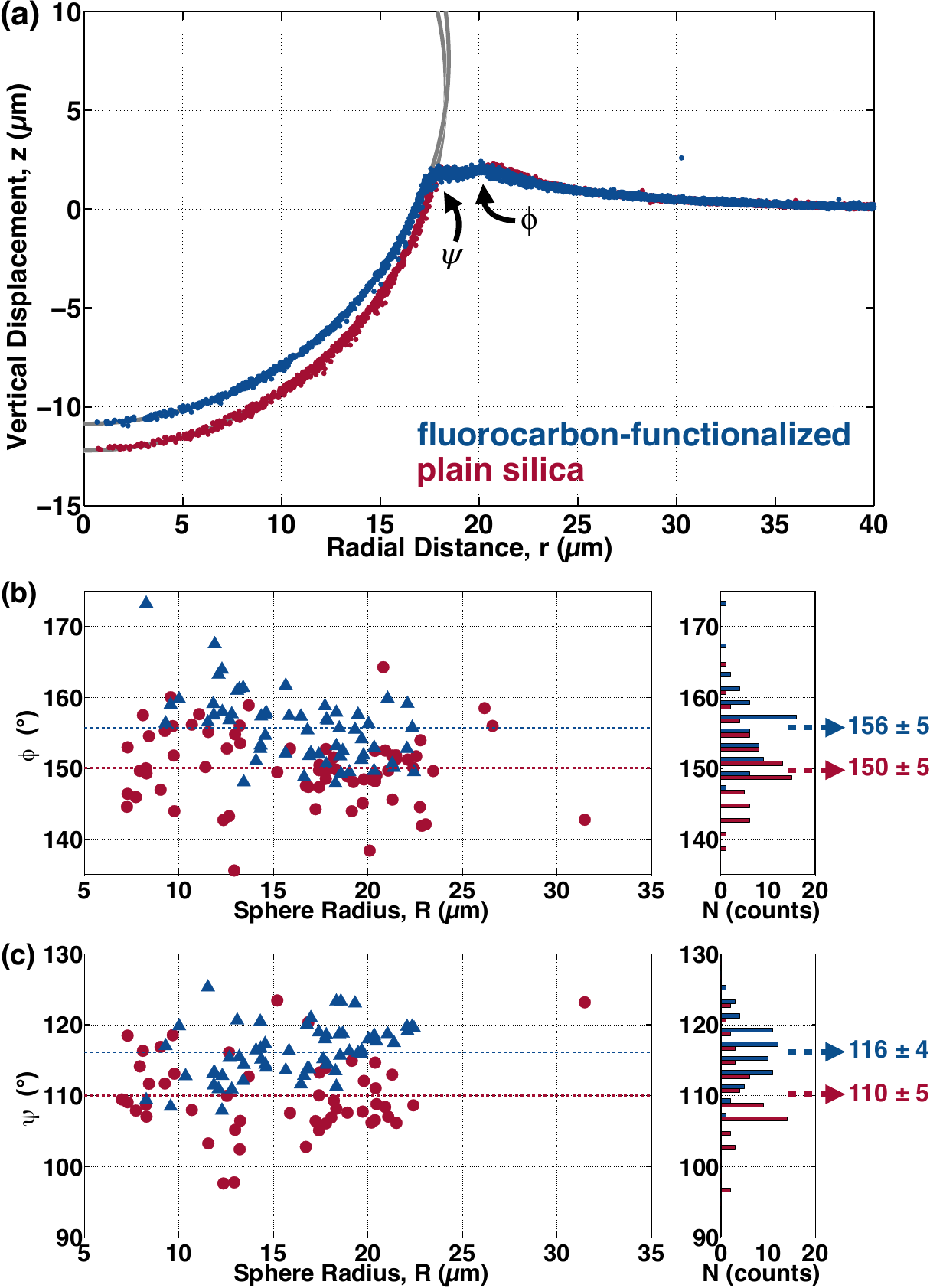}
\caption{\label{confocal} The structure of the gel's elastic network near contact. 
\emph{(a)} Confocal profiles of the surface of the silicone elastic network adhered to an 18.3-$\mu$m-radius plain silica sphere (red) and an 18.5-$\mu$m-radius  fluorocarbon-functionalized sphere (blue).
\emph{(b)} Contact angle, $\phi$, made by the elastic network as it abruptly changes direction during approach to contact.
\emph{(c)} Contact angle, $\psi$, made by the elastic network as it contacts the sphere. 
Both $\phi$ and $\psi$ are plotted versus sphere radius for both the fluorocarbon-functionalized (blue triangles) and plain silica (red circles) spheres. 
Dashed lines indicate the mean contact angle. 
Histograms of the measured contact angle are shown at right, with mean and standard deviation indicated.}
\end{figure}

To quantify the deformation of the gel's elastic network, we embed fluorescent tracers in the elastic network at the surface of the gel and image them using confocal microscopy.
For this experiment, we prepare flat $\sim$120-$\mu$m-thick silicone substrates on glass coverslips by spin-coating.
After curing, we adsorb 48-nm-diameter fluorescent spheres (Life Technologies, F-8795) from an aqueous suspension onto the PDMS. 
This procedure is identical to that described in \cite{TFMpaper} except that we do not chemically modify the silicone surface.
Then, following the procedure of Ref. \cite{StyleNatComm2013}, we sprinkle silica spheres onto the substrates and map the surface of the deformed elastic network by locating the fluorescent makers in 3D from confocal microscope images \cite{Gao2009}.
Examples of azimuthally-collapsed deformation profiles for each type of sphere are shown in Figure \ref{confocal}(a). 
We find that the dependence of indentation depth on particle size is consistent with our earlier study of the transition from elastic-dominated to capillary-dominated adhesion \cite{StyleNatComm2013}.

As expected, the elastic network rises gradually toward contact from the far field and conforms to the surface of the spheres underneath the particles.
However, the surface  of elastic network in the contact zone (Figure \ref{confocal}(a)) looks nothing like the free surface of the substrate (Figure \ref{brightfield}).
Specifically, the elastic network does not rise smoothly to contact the sphere with the expected contact angle and curvature.
Instead, it has a kink of angle $\phi$ a few microns from the sphere.
Eventually the elastic network comes into contact with the sphere with an angle $\psi$ well below the expected contact point.
A series of control experiments 
ruled out the possibility that the discrepancies between the structure of the contact zone in the brightfield and confocal experiments could be due to imaging artifacts.
Just like the contact angle of the free surface $\theta$ (Figure \ref{brightfield}(e)), the angles $\phi$ and $\psi$ are independent of sphere radius, as shown in Figure \ref{confocal}(b)(c).

\section{Adhesion-induced phase-separation}
\begin{figure}[h]
\includegraphics[width = 0.45 \textwidth]{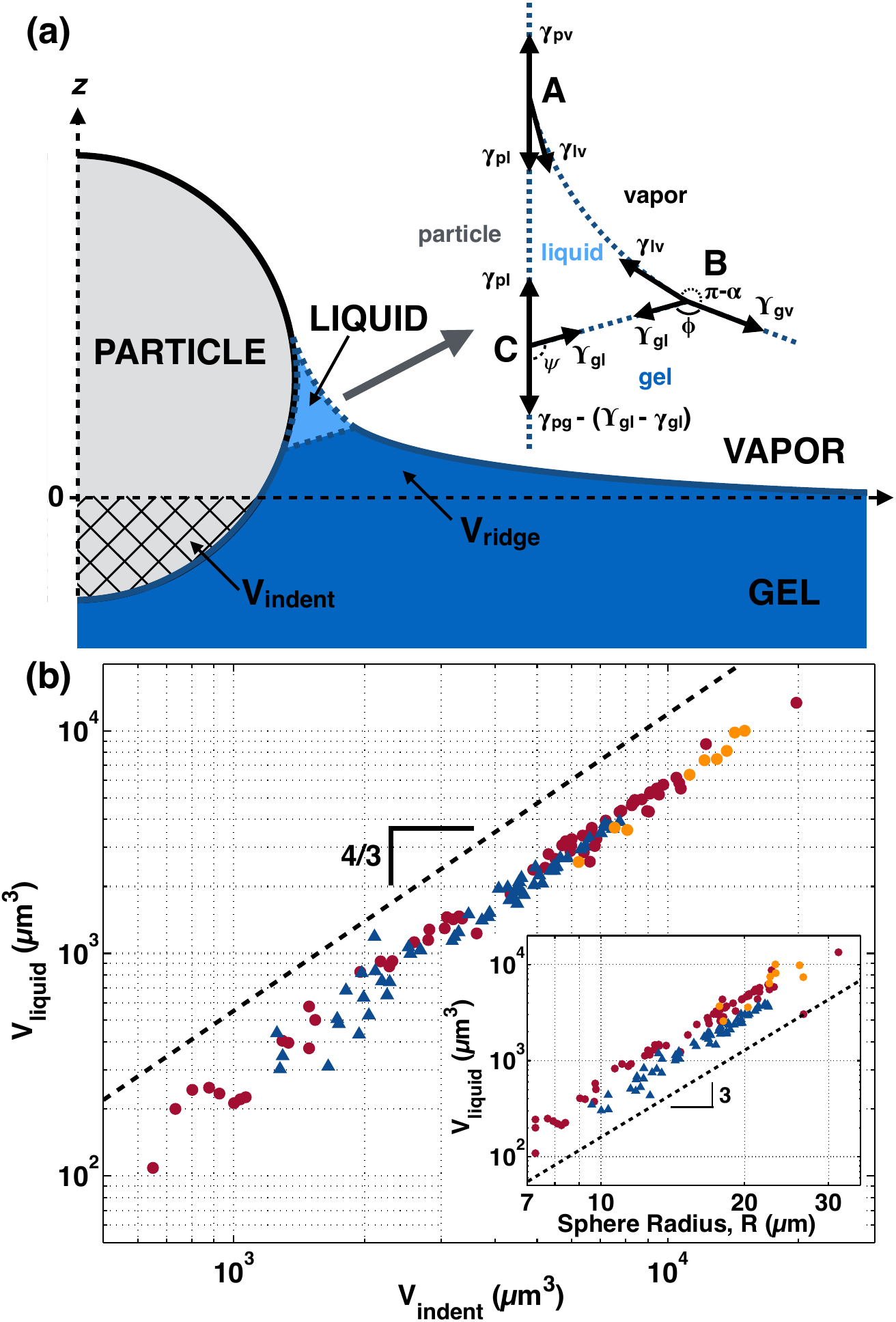}
\caption{\label{schematic} The structure and size of the four-phase contact zone. 
\emph{(a)} Schematic of the four-phase contact zone.
Inset: Schematic of the surface tension balance at each of the contact lines \textbf{A}, \textbf{B}, and \textbf{C}.
\emph{(b)} Plot of the volume of phase-separated liquid, $V_{\text{liquid}} = (V_{\text{indent}} - V_{\text{ridge}})$, vs. indented volume, $V_{\text{indent}}$, measured by integrating the confocal profiles. 
The data for plain spheres in air are plotted as red circles, for fluorocarbon-functionalized spheres in air as blue triangles, and for plain spheres under glycerol as orange circles. 
A dashed line of slope 4/3 is shown as a guide to the eye.
Inset: The same data plotted vs. sphere radius, with a dashed line of slope 3.}
\end{figure}

Comparison of the bright-field images in Figure \ref{brightfield}(a)(b) with the confocal images in Figure \ref{confocal}(a) suggests that liquid PDMS fills the space between the elastic network and the free surface, as shown schematically in Figure \ref{schematic}(a).  
In this way, the fluid can satisfy the Young-Dupr\'{e} wetting condition while the elastic network avoids an elastic singularity.
This adhesion-induced phase separation makes the zone of adhesive contact between a soft gel and a rigid object more complex than in adhesion to stiffer single-phase solids.
Instead of a single three-phase contact line, phase separation creates a four-phase contact zone in which air, silica, silicone liquid, and  silicone gel meet, as shown in Figure \ref{schematic}(a).
In addition to the standard contact line at \textbf{A}, the confocal experiments reveal two additional contact lines at \textbf{B} and \textbf{C}. 
The existence of particle-size-independent contact angles $\phi$ and $\psi$ at these contact lines strongly indicates that their geometry is governed by surface stresses and/or surface energies, as indicated in the Figure \ref{schematic}(a) inset.
The contact line at \textbf{A} is a conventional rigid solid-liquid-vapor contact line which satisfies the Young-Dupr\'{e} relation, as discussed above. 
The contact line at \textbf{B} follows a Neumann triangle construction at this soft solid-liquid-vapor contact line, as in Ref. \cite{StylePRL2013}. 
Finally, we expect the contact line at \textbf{C} to be described by a modified Young-Dupr\'e relation for a soft solid in contact with a rigid solid, as in Ref. \cite{StyleNatComm2013}. 

The structures of the contact lines at \textbf{B} and \textbf{C} therefore provide  information about the relevant surface stresses and surface energies \cite{cammarata1994surface}.  
For an ideal gel \cite {Hui2013}, the liquid phase dominates and the surface stress and surface energy of the the gel are identical and equal to the surface tension of the solvent \cite{Shuttleworth1950,Hui2013}.
However, recent measurements of the surface stress of gels have sometimes differed significantly from the surface tension of their fluid phase \cite{Nadermann2013,Chakrabarti2013, StyleNatComm2013, StylePRL2013}.
If our silicone gel were ideal, we would expect the surface of the gel to be equivalent to the surface of its solvent, such that $\Upsilon_{gl}=0.$
In that case, there would be no constraint on the contact angles, $\psi$ or $\phi$.
However, the existence of well-defined, size-independent values of $\psi$ and $\phi$, implies that $\Upsilon_{gl}>0$. 
Furthermore, we observe that $\psi > 90^\circ$, implying that $\gamma_{pg}>\gamma_{pl}$.
This means that the particle has a preference for making contact with the pure liquid over the gel.
This preference is only slightly changed by fluorocarbon-functionalization of the particle surface.

At contact line \textbf{B}, the surface tension of the liquid $\gamma_{lv}$ must balance the surface stresses of the gel, $\Upsilon_{gl}$ and $\Upsilon_{gv}$,  through the Neumann construction.  
In order to fully determine all the surface tensions, we also need to measure the difference in angle between the gel and the liquid free surfaces, $\alpha$, as indicated in the Figure \ref{schematic}(a) inset.  
In principle, $\alpha$ should be measurable as a discontinuity in the free surface at \textbf{B}.
However, our bright-field images do not reveal such a discontinuity (see Figure \ref{brightfield}).
This suggests that the angle $\alpha$ is small and cannot be resolved due to diffraction effects (as seen in Figure \ref{brightfield}(d)).
Small values of $\alpha$ are expected when $\Upsilon_{gl}$ and/or $(\Upsilon_{gv} - \gamma_{lv})$ are small.
Simplifying the Neumann condition for $\Upsilon_{gl} / \gamma_{lv} \ll 1$, we obtain $\alpha = (\Upsilon_{gl}/\gamma_{lv}) \sin{\phi} \ll 1$.
Further, by expanding both the horizontal and vertical force balances at \textbf{B} for $\epsilon = (\Upsilon_{gv} - \gamma_{lv})/\gamma_{lv} \ll 1$, we find that $1 + \epsilon = \cos{\alpha} - \sin{\alpha} \cot{\phi}$, which also results in small values of $\alpha$ for small $\epsilon$.

Although we cannot measure $\alpha$ directly in these experiments, we can put a rough upper bound on its magnitude by combining our bright field and confocal results for the geometry of the contact zone.
These observations allow us to constrain $\alpha$ between 0$^\circ$ and  10$^\circ$.
This bounds the values of the solid surface stresses such that $0 < \Upsilon_{gl} \lesssim 0.4 \gamma_{lv}$ and $\gamma_{lv} < \Upsilon_{gv} \lesssim 1.3 \gamma_{lv}$.
More precise measurements of the free surface profile at contact line \textbf{B} will be required for precise measurement of the solid surface stresses.

Surface stresses and energies fix the geometries of the corners of the phase-separated liquid region at \textbf{A}, \textbf{B}, and \textbf{C}, but this is not sufficient to determine its overall size, $V_{\text{liquid}}$.
Since the liquid is incompressible but the elastic network is not \cite{Geissler1980, SuoPoisson2010}, $V_{\text{liquid}}$ must equal the change in volume of the elastic network due to the adhesion of the sphere.
We define $V_{\text{indent}}$ as the volume occupied by the sphere below the plane of the undeformed silicone surface, and $V_{\text{ridge}}$ as the volume of the elastic network displaced above the undeformed surface, as indicated in Figure \ref{schematic}(a).
Thus, we can measure $V_{\text{liquid}}$ from our confocal profiles as $V_{\text{indent}}-V_{\text{ridge}}$.
We compute these volumes by numerical integration of the axisymmetric confocal profiles.  

We plot $V_{\text{liquid}}$ vs. sphere radius in the inset of Figure \ref{schematic}(b). 
We see that the dependence of $V_{\text{liquid}}$ on sphere size differs for the different surface functionalizations, but scales approximately as $R^3$.
This suggests that $V_{\text{liquid}}$ may be related to volume, rather than surface effects.
We find that all of the data collapses if we instead plot $V_{\text{liquid}}$ versus $V_{\text{indent}}$, as shown in the main panel of Figure \ref{schematic}(b).  
The volume of the phase-separated contact zone scales as a power-law with exponent 4/3 over this range of indentation volumes.
The more the elastic network is compressed by the spontaneous indentation of the particle, the larger the volume of incompressible liquid that phase separates from the elastic network.
This collapse is robust not only for the fluorocarbon-functionalized and plain silica spheres, but also after changing the balance of surface energies by covering the sphere and substrate with glycerol. 
It can even work when the system is out of equilibrium, as some of the glyercol-covered data points were not given enough time to  equilibrate to their new indentation depth.
Dimensionally, the prefactor for this power-law collapse must have dimensions of 1/[length].
Fitting  to $V_{\text{liquid}} = (1/L') V_{\text{indent}}^{4/3}$, we measure $L' = 38$ $\mu$m, which is about ten times the elastocapillary length. 

We have seen that during adhesion with a rigid object, a compliant gel phase-separates near the contact line to create a four-phase contact zone with three distinct contact lines.
The total volume of the phase-separated region is set by the extent of indentation and the compressibility of the gel's elastic network.
The geometries of these contact lines are independent of the size of the particles and suggest that the gel-vapor solid surface stress, $\Upsilon_{gv}$, and the liquid-vapor surface tension, $\gamma_{lv}$, are  different, and that the solid surface stress between the gel and the liquid, $\Upsilon_{gl}$, is nonzero.

These findings substantially change our understanding of the mechanics of adhesion with gels.
This new understanding of the geometry of contact and the balance of forces at work modifies both future theoretical work and engineering design of soft interfaces.
While in many situations, a gel can be considered a single, homogenous material, our results demonstrate that under extreme or singular conditions -- such as at a contact line -- the nature of a gel as a multi-phase material becomes critically important.
This may have important implications not just for silicone materials, but also for materials like hydrogels, which have recently been the subject of significant research efforts \cite{Mooney2001, Mooney2003, Suo2012}. 
Because hydrogels can easily be much more compressible than the silicone gels studied here \cite{Geissler1980, SuoPoisson2010}, it is possible that they will be even more susceptible to phase separation during contact.

We thank Manjari Randeria and Ross Bauer for help with sample preparation, and Dominic Vella for useful discussions and for help with the MATLAB code for the constant curvature analyses. 
We acknowledge funding from the National Science Foundation (CBET-1236086). R.W.S. also received funding from the John Fell Oxford University Press (OUP) Research Fund.


\end{document}